\documentclass{aa}  
\usepackage{graphicx}
\usepackage[]{natbib}
\usepackage{color}
\bibpunct{(}{)}{;}{a}{}{,} 
\usepackage{txfonts}
\begin{document}

   \title{Synchrotron radiation and absence of linear polarization in the colliding wind binary WR 146}

  \author{C. A. Hales\inst{1}, P. Benaglia\inst{2,3}, S. del Palacio\inst{2,3}\thanks{Fellow of CONICET}, G. E. Romero\inst{2,3},
          \and B. S. Koribalski\inst{4}}

   \institute{
            National Radio Astronomy Observatory, P.O. Box 0, Socorro, NM 87801, USA; \email{chales@nrao.edu}
       \and Instituto Argentino de Radioastronom\'{\i}a, CCT-La Plata, CONICET,
            Argentina
       \and Facultad de Ciencias Astronomicas y Geofisicas, Universidad Nacional de La Plata, Pasea del Bosque s/n, 
            1900 La Plata, Argentina
       \and Australia Telescope National Facility, CSIRO Astronomy \& Space Science
            P.O. Box 76, Epping, NSW 1710, Australia
        }

   \date{Received Month Day, Year; accepted Month Day, Year}

  \abstract
   {Several massive early-type binaries exhibit non-thermal emission which has been attributed to synchrotron radiation from 
   particles accelerated by diffusive shock acceleration (DSA) in the wind-collision region (WCR). If the magnetic field in the 
   strong shocks is ordered, its component parallel to the shock front should be enhanced, and the resultant synchrotron radiation
   would be polarized. However, such polarization has never been measured.}
   {We aim to determine the percentage of linearly polarized emission from the well-known non-thermal radio emitter WR~146, a WC6+O8 system.}
   {We performed spatially-unresolved radio continuum observations of WR~146 at 5~cm and 20~cm with the Karl G. Jansky Very Large Array.
   We constructed a numerical model to investigate a scenario where particles are accelerated by turbulent magnetic reconnection (MR),
   and we performed a quantitative analysis of possible depolarization effects.}
   {No linearly polarized radio emission was detected. The data constrain the fractional linear polarization to less
   than 0.6\% between 1 to 8 GHz. This is compatible with a high level of turbulence and a dominant random component
   in the magnetic field. In this case the relativistic particles could be produced by turbulent magnetic reconnection.
   In order for this scenario to satisfy the required non-thermal energy budget, the strength of the
   magnetic field in the WCR must be as high as $\sim 150$ mG. However, if the magnetic field is ordered and DSA is ongoing,
   then a combination of internal and external Faraday rotation could equally account for the depolarization of the emission.}
   {The absence of polarization could be caused by a highly turbulent magnetic field, other depolarization mechanisms such as
   Faraday rotation in the stellar wind, or a combination of these processes. It is not clear whether it is possible to develop the
   high level of turbulence and strong magnetic fields required for efficient MR in a long-period binary such as WR~146.
   This scenario might also have trouble explaining the low-frequency cutoff in the spectrum. We therefore favor a scenario
   where particles are accelerated through DSA and the depolarization is produced by mechanisms other than a large ratio
   between random to regular magnetic fields.}

   \keywords{Polarization -- Radio continuum: stars -- Radiation mechanisms: non-thermal -- Stars: individual: WR 146 -- Stars: winds, outflows}

\titlerunning{Absence of linearly polarized emission from WR~146}
\authorrunning{Hales et al.}

   \maketitle
%

\section{Introduction} \label{Sec:Introduction}

Hot, early-type stars (O-B2, Wolf-Rayet) are commonly found in stellar systems. If the two components of the system 
exhibit strong winds, these can interact forming a wind collision region (WCR); we refer to such systems as
colliding-wind binaries (CWBs). Colliding stellar winds are clearly detected at a few GHz using radio 
interferometers, as methodical studies such as \cite{Bieging1989} show. Observations of these systems
over a range of radio frequencies show flux density decreasing with the observing frequency, 
something characteristic of non-thermal (NT) radiation. The recent catalog of particle-accelerating colliding-wind 
binaries (hereafter PACWB) compiled by \cite{DeBecker2013} presents very detailed information of what is known
about these systems. Their emission is interpreted as having a synchrotron origin (i.e. due to relativistic
electrons moving in the presence of a magnetic field). Synchrotron radiation is expected to be polarized.
However, to date there are no reports of polarized radio emission from a PACWB.

According to the PACWB catalog there are more than 40 known binary systems capable of producing synchrotron 
radiation. One of these is WR~146, the brightest WR star at radio wavelengths and the fourth to be
associated with a NT radio source \citep{Felli1991}. It is located at a distance of~1.2 kpc and is 
thought to be a WC6+O8 system \citep{Niemela1998}. MERLIN observations at 5~GHz \citep{Dougherty1996} show a NT
source located north of the primary star. The system was observed in 2004 with the VLA in combination with
the VLBA Pie Town antenna, from 1.4 to 43~GHz \citep{OConnor2005}. The 43~GHz observations
reveal two components, separated by about 0.2~arcsec: a northern source coincident with the secondary star and a southern
source coincident with the primary star. The 1.4~GHz observations reveal that the NT source is located part-way
between the two components, close to the northern one. Only the NT source is detected at the lower frequencies
(1.4~GHz, 5~GHz). The system was monitored over 10~yr with the WSRT at 350~MHz and 1.4~GHz. Variability with
period 3.38~yr suggests the presence of a third component that modulates the O star wind \citep{Setia2000}.

The polarization characteristics of a synchrotron source can potentially provide useful and unique information about physical conditions 
within the source \citep{Cioffi1980}. An accurate determination of the polarization degree (PD) allows one, in principle, to
distinguish among different scenarios.
If the observed PD is between a few percent up to a few tens, this would provide decisive evidence 
of the synchrotron origin of the NT emission and constrain the degree of turbulence in the plasma. 
If an extremely high polarization (above 70\%) is detected, the underlying physics of particle acceleration 
in CWBs will be shown to be different from what has been traditionally assumed in most theoretical models.
Finally, if no polarization is detected (i.e., PD$\lesssim 1\%$), the main 
acceleration and emission mechanisms of the standard model of CWBs
-- diffusive shock acceleration (DSA) and synchrotron radiation, respectively -- might need to be reconsidered.
Therefore, polarization measurements can improve our knowledge of CWBs. 

This work presents the first radio polarimetric study of a CWB, focusing on WR 146. We describe our
observations and data reduction in Sec.~\ref{Sec:Observations}. Results are presented in Sec.~\ref{Sec:Results}.
In Sec.~\ref{Sec:Discussion} we discuss acceleration processes and interpret our polarization findings.
Concluding remarks and directions for future research are presented in Sec.~\ref{Sec:Conclusions}.


\section{Observations and Data Reduction} \label{Sec:Observations}

We observed WR 146 with the Karl G. Jansky Very Large Array (VLA) on 2015 May 4 under Project Code 15A-480.
The array was in B configuration. Full polarization observations were performed in two frequency bands.
These were centered at 1.5~GHz (20~cm, L-band) and 6~GHz (5~cm, C-band) spanning bandwidths of 1~GHz and 4~GHz,
respectively. In the 20~cm band, the correlator was configured to deliver 16 spectral windows with $64\times1$~MHz
channels. The 5~cm band was configured to deliver 32 spectral windows with $64\times2$~MHz channels. The time
sampling was 3~s for both bands. The total observing time on WR~146 was 19~min at L-band and 14~min at C-band.
We observed 3C138 for the purpose of flux density and position angle calibration, J2007+4029 for amplitude and phase
calibration, and the unpolarized source J2355+4950 for instrumental leakage calibration. J2355+4950 is less than
0.25\% polarized at L-band and C-band.

The data were reduced using version 4.3.0 of the CASA package \citep{2007ASPC..376..127M}.
Hanning smoothing was performed. RFI was identified manually and conservatively flagged.
The flux density scale was referenced to the most recent 2012 value for 3C138 from \citet{2013ApJS..204...19P}.
The position angle of 3C138 was assumed to be $-11^\circ$ on all spatial scales across L-band and C-band.
This is within a few degrees of the values presented by \citet{2013ApJS..206...16P} based on D configuration
observations. However, their values are only loosely appropriate for translation to our B configuration data
because 3C138 exhibits polarization structure on sub-arcsecond scales \citep{1997A&A...325..493C}. We conservatively
estimate that any intrinsic rotation measure (RM) left unaccounted in our model of 3C138 is no worse than 2~rad~m$^{-2}$.
No dedicated corrections were performed to account for atmospheric Faraday rotation taking place within the
ionosphere and plasmasphere (the necessary CASA functionality contained an error in version 4.3.0, so it was not used).
Line of sight atmospheric RMs throughout our observation were approximately constant at $\sim6$~rad~m$^{-2}$,
estimated by combining GPS-derived total electron content data from the International GNSS Service with the International
Geomagnetic Reference Field model of the Earth's magnetic field. Without correction, this atmospheric
RM was absorbed into our position angle calibration, imparting a residual $\sim1$~rad~m$^{-2}$
on our target data. Combining these two uncertainties, we estimate that our WR 146 data are affected by a
systematic RM of no more than approximately 2~rad~m$^{-2}$. This is negligible given our resolution in
Faraday depth space, described below. We performed one round of phase self-calibration on WR 146 in each band.

The total intensity data were imaged per band as well as per spectral window,
the latter every 64~MHz in the 20~cm band (minus 3 windows affected by RFI) and every 128~MHz
in the 5~cm band (minus 4 windows affected by RFI). The continuum rms noises in the full-band 20~cm and 5~cm
images are 40~$\mu$Jy~beam$^{-1}$ and 12~$\mu$Jy~beam$^{-1}$, respectively. These values are worse than
theoretical expectations (almost double) due to unusually disruptive RFI. However, taking into account
observational contingencies, the sensitivities are sufficient for the objectives in this work. The spatial
resolutions in the full-band 20~cm and 5~cm images are $\sim4"$ and $\sim1"$, respectively. Flux density measurements
in the full-band and per spectral window images were performed using {\tt BLOBCAT} \citep{2012MNRAS.425..979H}.

We imaged the Stokes Q and U data by pre-averaging every 2 input spectral channels to form an output
image cube with 2~MHz channels across L-band and 4~MHz channels across C-band. Each image plane was then
smoothed to a common spatial resolution of $14"\times6"$, set by the lowest frequency. Using this data
we performed RM synthesis \citep{2005A&A...441.1217B}. We examined the data in four different frequency ranges:
the combined 1--2~GHz and 4--8~GHz data, and subsets 1--2~GHz, 4--8~GHz, and 6--8~GHz. The resolution
in Faraday depth space for each of these frequency ranges is approximately $40$~rad~m$^{-2}$, $50$~rad~m$^{-2}$,
$900$~rad~m$^{-2}$, and $3000$~rad~m$^{-2}$, respectively. The effective rms noise
\citep[$\sigma_{QU}$;][]{2012MNRAS.424.2160H} for each of the respective ranges is 22~$\mu$Jy~beam$^{-1}$,
30~$\mu$Jy~beam$^{-1}$, 34~$\mu$Jy~beam$^{-1}$, and 51~$\mu$Jy~beam$^{-1}$.


\section{Results} \label{Sec:Results}

We detect WR 146 as a single spatially-unresolved source in the 20~cm and 5~cm total intensity images.
The flux density is $79.8 \pm 1.6$~mJy at 1.52~GHz and $35.7 \pm 0.7$~mJy at 6.00~GHz, obtained
from the full-band images. The errors include a systematic absolute flux density uncertainty of 2\%
added in quadrature to the statistical errors previously quoted.

The spectrum obtained from the per spectral window data is displayed in Fig.~\ref{fig:fit_alpha} and tabulated in the Appendix.
The figure shows marginal evidence for a turnover below 1.5~GHz. The spectral index ($\alpha$ such that
$S_\nu \propto \nu^{\alpha}$) is $-0.58 \pm 0.02$ when fit to the flux densities from all spectral windows. It is
possible that the flux densities at the lower frequencies are affected by absorption. To account for this, we fit
only the higher frequency data from spectral windows within the 5~cm band, obtaining a softer value of $-0.62 \pm 0.02$
(Fig.~\ref{fig:fit_alpha}). To highlight the putative turnover, in Fig.~\ref{fig:fit_alpha} (and later in
Fig.~\ref{fig:spectrum}) we also plot measurements at 327~MHz from Benaglia \& Ishwara-Chandra (private communication;
observed with GMRT in Oct 2014) and \citet{Taylor1996}, where the cutoff becomes more pronounced. Note,
however, that their observations are not contemporaneous with those presented in this work, so their
data points are not expected to correspond with our spectral fit. 

 \begin{figure}
   \centering
   \includegraphics[width=0.3\textwidth,angle=270]{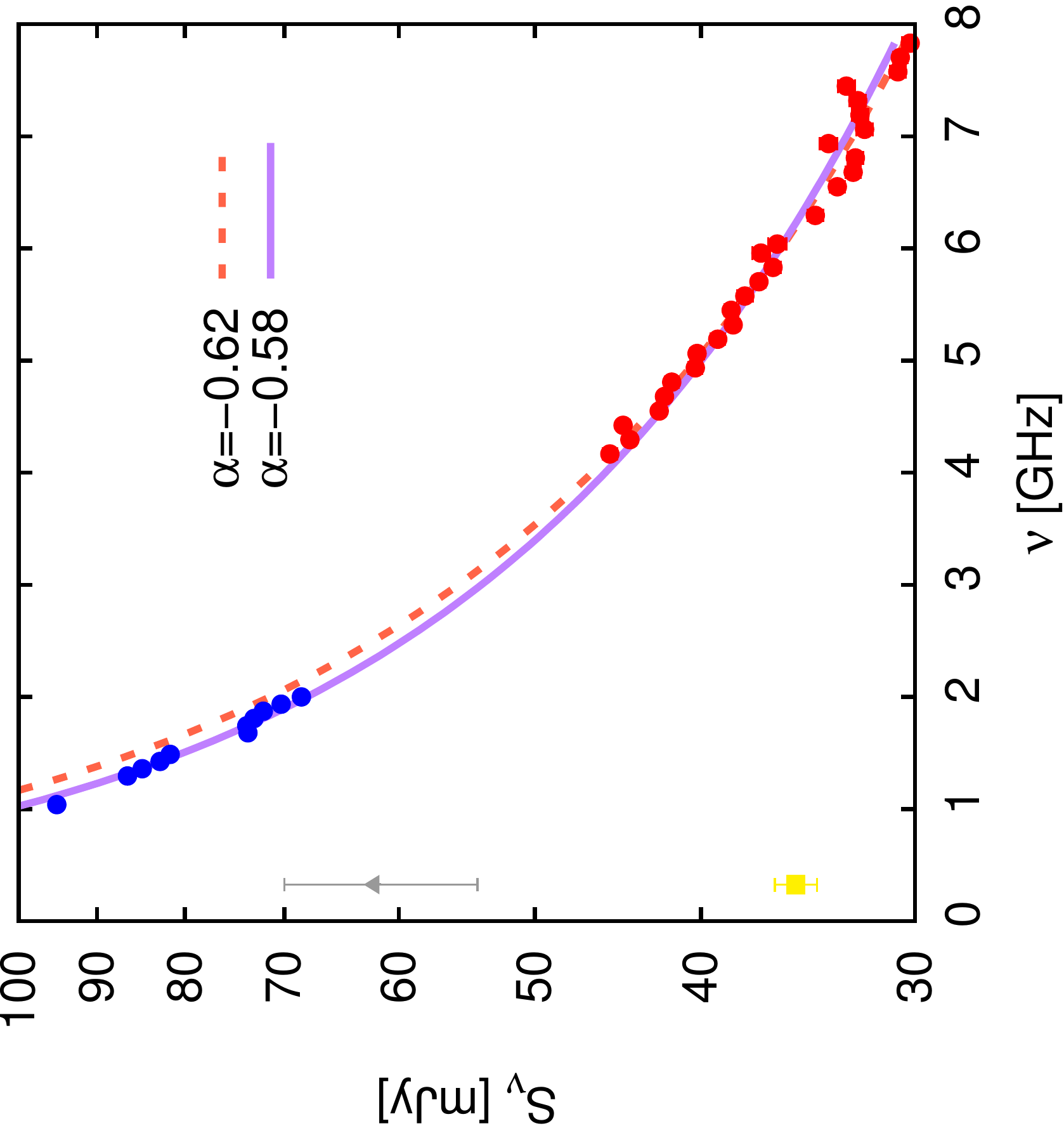}
   \caption{Radio observations at L-band (blue points) and C-band (red points).
   We show the spectral fit $S_\nu \propto \nu^{\alpha}$ to the whole 
   data set (solid purple line) and only to the C-band data (dashed orange line).
   We also show the flux densities measured at 327 MHz by \cite{Taylor1996}
   (grey triangle) and Benaglia \& Ishwara-Chandra (private communication, observed
   in Oct 2014, yellow square).}
   \label{fig:fit_alpha}
 \end{figure}

We do not detect any linearly polarized emission from WR 146 within the 4 frequency ranges examined.
Upper bounds\footnote{Adhering to the terminology regarding bound versus limit from \citet{2010ApJ...719..900K}.}
are reported in Table~\ref{tab:results}.
\begin{table}
  \centering	
  \begin{tabular}{cccc}
   \hline	
   \textbf{Data}	& \textbf{FD range}			& \textbf{LPFD} \\
   \hline
    1--2, 4--8~GHz	& $\pm 2 \times 10^6 $rad m$^{-2}$	& $<140$~$\mu$Jy \\
    1--2~GHz		& $\pm 4 \times 10^4$~rad~m$^{-2}$ 	& $<120$~$\mu$Jy \\
    4--8~GHz		& $\pm 10^6$~rad~m$^{-2}$		& $<180$~$\mu$Jy \\
    6--8 GHz		& $\pm 10^6$~rad~m$^{-2}$		& $<190$~$\mu$Jy \\
   \hline
  \end{tabular}
  \caption{Upper bounds on the linearly polarized flux density (LPFD)
           for different portions of the data. The bounds give the
	   peak in the Faraday depth (FD) spectrum over the FD range
	   examined using RM synthesis.
	   }
  \label{tab:results}
\end{table} 
In each case, the statistical significance is below the Gaussian equivalent of a 4 sigma
detection \citep{2012MNRAS.424.2160H}. Combining these results with total intensity 
flux densities, the PD at L-band is less than 0.15\% ($<120\,\mu$Jy/80~mJy), at C-band
it is less than 0.5\% ($<180 \,\mu$Jy/36~mJy), and within the upper half of C band (6--8~GHz)
it is less than 0.6\% ($<190 \,\mu$Jy/32~mJy). Thus overall, the WR~146 emission exhibits less
than 0.6\% fractional polarization across 1-8~GHz.


\section{Discussion} \label{Sec:Discussion}

Since the pioneering studies of \cite{Eichler1993}, the NT radio emission of PACWBs has been explained as synchrotron 
radiation produced by a population of relativistic electrons interacting with the local magnetic field. It has been suggested 
that the most suitable mechanism to account for the acceleration of electrons in PACWBs is DSA through a first-order Fermi process 
\citep{Romero1999, Benaglia2003, Pittard2006b, Reitberger2014a}, although magnetic reconnection (MR) is also 
possible \citep{Jardine1996}. For DSA to operate
efficiently, a strong shock, a turbulent motion of the plasma, and a disordered magnetic field are required. The plasma is 
expected to develop instabilities near the contact discontinuity \citep{Stevens1992}, which would provide the required 
turbulence for this scenario, although a very high turbulence might debilitate the shock and therefore
reduce the efficiency of DSA. On the other hand, for MR to be efficient, small-scale turbulent motions are necessary \citep{Lazarian1999}. 
Hence, this mechanism cannot be discarded in a very turbulent regime \citep{Falceta2012}. 
The magnetic field is not required to be high for DSA \citep{Bosch-Ramon2012}, in contrast to the case of MR.\\

In any case, the high-energy electrons are expected to obtain a power-law distribution in
energy with a spectral index $p \sim 2$. The resultant synchrotron radiation is intrinsically polarized up to a value of 
$\Pi_i(p) \sim 70\%$. However, a turbulent magnetic field greatly decreases the PD. 
If the random component of the field is $B_\mathrm{r}$ and the ordered field is $B_0$, the observed PD becomes \citep{Burn1966}:

\begin{equation}
 \Pi_{\mathrm{obs}}(p)= \Pi_\mathrm{i}(p) \, \frac{B_0^2}{B_0^2+B_\mathrm{r}^2} \, \xi(\lambda),
\end{equation}
where the function $\xi(\lambda)$ takes into account possible depolarization effects (wavelength-dependent in some cases; Sect.~\ref{sec:depol}).
Since the magnetic field lines are frozen in the plasma, the random component of the magnetic field is linked to the degree of 
turbulence in the WCR, measured by the ratio $B_\mathrm{r}/B_0$. Therefore, the PD is strictly related to the 
degree of turbulence, and a small polarization could be indirect evidence of a highly turbulent medium. Considering solely this effect, 
our result of non-detection of polarization in WR~146 points to a very high ratio $B_\mathrm{r}/B_0 > 8$. 
DSA is not expected to be efficient under such conditions due to the consequently weaker shocks, which leads to the conclusion
that MR could be the principal acceleration mechanism acting on this PACWB.

\subsection{Turbulent Magnetic Reconnection}
MR is a process that develops when two converging flows meet. If the magnetic fields carried by the flows have 
opposite polarity, the magnetic field lines reconnect and magnetic energy is released. This energy can be converted to kinetic 
energy, i.e., heating and/or particle acceleration, in multiple small-scale MR events that develop in turbulent plasmas. 
In such case particles are 
accelerated in multiple magnetic islands \citep{Kowal2011}. The acceleration timescale is $t_{\mathrm{acc}}^{-1} = \eta_\mathrm{acc} e c B/E$, 
where $\eta_\mathrm{acc} \sim 0.3 (v_{\mathrm{rec}}/c)^2$ is the acceleration efficiency, $v_{\mathrm{rec}}$ the 
reconnection velocity, $e$ and $E$ the charge and energy of the particle, respectively, and $B$ the ambient magnetic field. Depending on 
the conditions in the plasma, the spectral index
can be $p < 2.5$ \citep{Drury2012}, which is in agreement with the injection spectral index derived from our observations,
$p = -2\alpha +1 \simeq 2.3 \pm 0.1$.

We applied the NT emission model described in \cite{delPalacio2016} to test whether turbulent MR can account for the 
accelerated particles in WR~146. This model considers the adiabatic wind shocks to be thin enough to neglect their width, and that the 
relativistic particles are attached to the flow streamlines through the chaotic $B$-component. The NT particle content is modeled 
as sets of linear-emitters. In each linear-emitter, a population of relativistic particles is injected at a given location with a 
distribution $Q(E)\propto E^{-p}$, and evolve until reaching the stationary particle distribution at positions along 
the streamline. The relevant energy losses (IC scattering and synchrotron for electrons; 
p-p interactions for protons) and the advection of particles are taken into account when computing the particle evolution.
The adopted system parameters are summarized in Tab.~\ref{tab:parameters}. The adopted parameters yield
a linear distance of $D\sim200$~AU. The small value of $\eta$ means that the shocked structure is much closer to the secondary star.

  \begin{table}
    \centering
    \begin{tabular}{lcc}
    \hline
    \textbf{Parameter}		&	\textbf{Value}		&	\textbf{Unit}	\\
    \hline
    Primary spectral type	&       WC6			&			\\
    Secondary spectral type	&       O8			&			\\
    Distance			&       $d=1.2$			&       kpc		\\ 
    System separation		&	$D_\mathrm{proj}= 160$	&       mas		\\
    Inclination of the orbit	&	$i = 30$		&       $^\circ$	\\
    Period			&	$P = 3.38$		&       yr		\\    
    Wind momentum ratio		&	$\eta=0.1$		&                     	\\
    
    \hline
   $T_{\mathrm{eff}1}$ 		&	49000			&       K		\\
   $R_1$			&       1.5			&       R$_\odot$	\\    
   $v_{\infty_1} \qquad$ ~(Primary)&	2900			&       km s$^{-1}$	\\ 
   $\dot{M_1}$                  &	$5.5\times 10^{-5}$	&	M$_\odot$ yr$^{-1}$\\ 
   $V_{\mathrm{rot}_1}/v_{\infty_1}$&	$\sim 0.1$		&			\\ 
   $\mu_{\mathrm{w}1}$ 		&  	5.29       		&	               	\\ 
   $T_{\mathrm{w}1}$        	&	8000            	&      	K 		\\
    
    \hline
   $T_{\mathrm{eff}2}$     	&      	32000               	&       K              	\\
   $R_2$                     	& 	10                    	&       R$_\odot$      	\\    
   $v_{\infty_2} \qquad$ ~(Secondary)&	1600                	&       km s$^{-1}$    	\\ 
   $\dot{M_2}$                	&   	$8\times 10^{-6}$  	&       M$_\odot$ yr$^{-1}$\\ 
   $V_{\mathrm{rot}_2}/v_{\infty_2}$&	$\sim 0.1$           	&     			\\ 
   $\mu_{\mathrm{w}2}$    	&  	1.4           		&           	      	\\ 
   $T_{\mathrm{w}2}$       	&       $0.4 T_{\mathrm{eff}2}$&   	K             	\\

    \hline
    \end{tabular}
    \caption{Parameters of the WR~146 system for the primary (sub-index 1) and the secondary (sub-index 2). $\mu$ is the mean atomic weight
    and $v_\infty$ is the wind terminal velocity. For further information on these system parameters, refer to \cite{Setia2000} and \cite{Dougherty2000}.}
    \label{tab:parameters}
  \end{table}
  
The WCR consists of the shocked stellar wind from the primary (S1) and the shocked stellar wind from the secondary (S2),
each with different properties. Magnetic field notation is as follows: $B_{*1}$ and $B_{*2}$ are the stellar surface magnetic fields,
and $B_{1}$ and $B_{2}$ are the compressed fields in the shocks S1 and S2 which are (in our simple model) the WCR.
In both shocks we considered a reconnection velocity given by $v_{\mathrm{rec}} = 0.6 v_\mathrm{A}$ \citep{delValle2012}, where $v_\mathrm{A}$ 
is the Alfv\'en velocity. We also assumed that $\sim 4\%$ of the released magnetic energy goes to accelerating electrons
\citep[][and references therein]{delValle2011}.

\begin{figure}
   \centering
     \includegraphics[width=0.34\textwidth, angle=270]{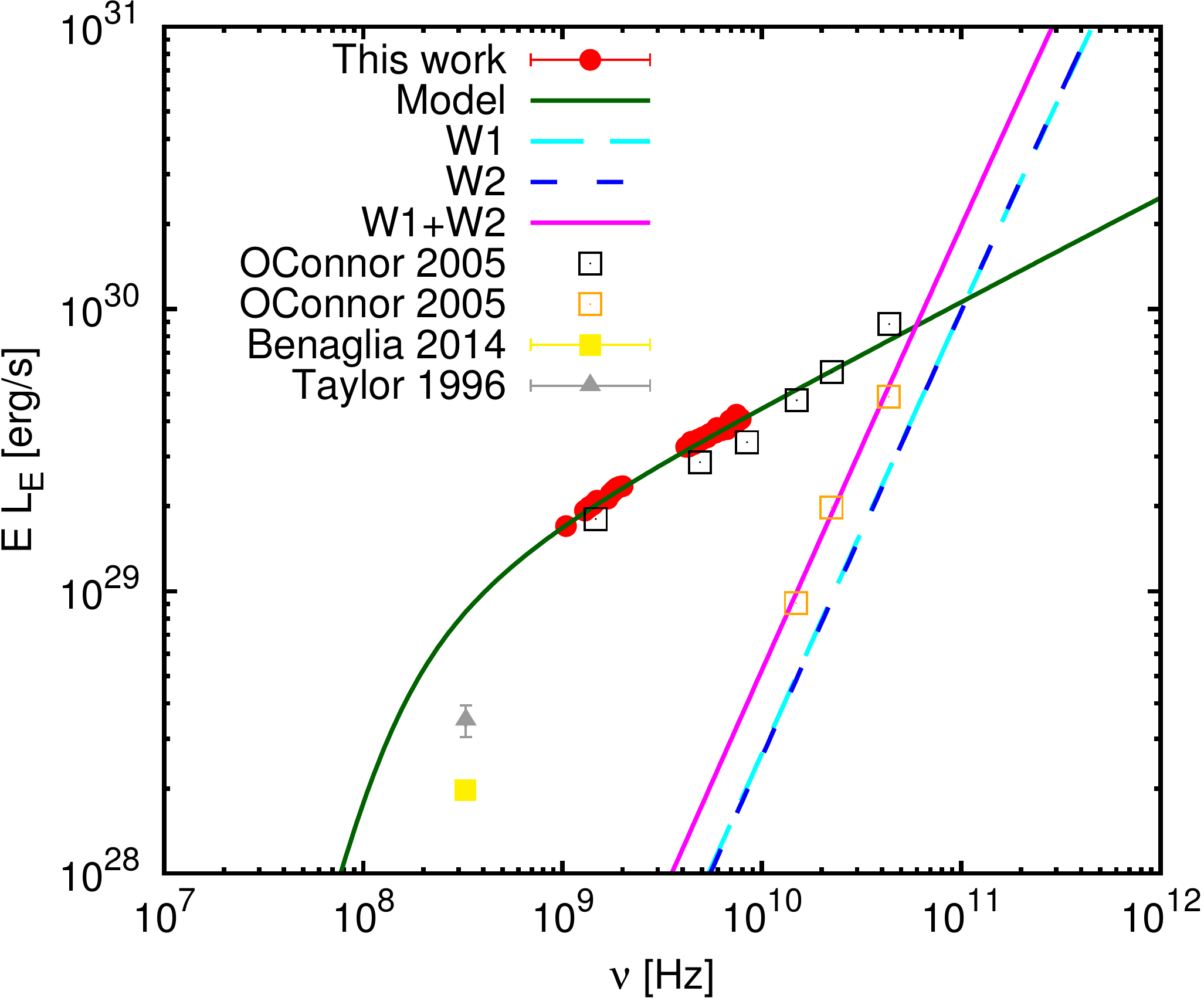}
      \caption{Luminosities in the 20~cm and 5~cm bands from our per spectral window data (red points).
      The solid green curve is the simulated emission considering acceleration by MR, and synchrotron emission suffering free-free absorption
      in the stellar wind. We also include data points from \cite{Taylor1996} (grey triangle) and Benaglia \& Ishwara-Chandra
      (private communication, observed in Oct 2014, yellow circle), and from \cite{OConnor2005} for the total flux density
      (dark grey squares) and the WR emission (orange squares). The 
      straight lines represent the separated (dashed) and summed (solid) free-free emission from the winds using Eq.~8 from \cite{Wright1975}.}
         \label{fig:spectrum}
   \end{figure}

Under the mentioned conditions, our results show that as long as the WR star is not extremely magnetized (surface magnetic field 
$B_{*1}< 2000$~G), or that no highly efficient $B$-amplification is taking place in S1, then the NT emission is 
completely dominated by S2 (i.e. $B_{2}$ is the only relevant magnetic field).
Fitting the observed energy fluxes requires that in S2 the magnetic field energy density is 38\% of the 
thermal energy density, which corresponds to a magnetic field intensity of $B_2 \sim 150$~mG, and $\eta_\mathrm{acc} \sim 10^{-7}$. 
For typical values of the Alfv\'en 
radius $r_\mathrm{A} = R_2$ and rotational speed $V_\mathrm{rot} = 250$~km~s$^{-1}$, this yields a maximum value for the stellar 
surface magnetic field $B_{*2} < 800$~G; this condition is relaxed if magnetic field amplification is taking place in the shocks.
Fig.~\ref{fig:spectrum} shows a good agreement between our model and the new radio observations. Features of the non-simultaneous observations 
are also reproduced qualitatively, although it is not clear if the cut-off at lower energies can be explained by free-free absorption 
in the stellar wind (suppression due to Razin-Tsytovich effect is negligible due to the high value of $B$; we further discuss this in 
Sect.~\ref{sec:variability}).

In the MR model, the acceleration time depends heavily on $B$. In S1 no efficient acceleration takes place, 
while in S2 electrons can reach energies up to $\sim 100$ GeV, this being the maximum energy limited by IC losses with stellar photons. 
The expected IC luminosity is not very high and falls below the gamma-ray detector sensitivities of the LAT
on board the \textit{Fermi} satellite (4-yr of integration time) and the Cherenkov Telescope Array (CTA, 50-h).

\subsection{Depolarization processes} \label{sec:depol}
Synchrotron radiation can be depolarized by various radiation transfer effects. In transparent radio sources
the main effect is Faraday rotation (FR). FR alters the plane of polarization and can reduce the observed PD.
\cite{Sokoloff1998} reviewed the different mechanisms that produce depolarization of synchrotron radio emission. In what follows we 
will analyze the most relevant of these processes in CWBs, besides the one presented above. To be more conservative,
we assume a classical DSA scenario. In such case, the available energy to accelerate NT particles is larger, as it is provided by 
the wind kinetic energy instead of the magnetic field. Moreover, the acceleration timescale in the fast shocks is shorter than in the MR 
scenario. To fit the observed synchrotron fluxes, a value of $B_2 \sim 40$~mG is then sufficient.

\begin{enumerate}

 \item {Reduced polarization due to thermal fraction: A significant contribution from thermal (unpolarized) emission can
 reduce the observed PD. However, the thermal contribution in L-band is below 1\% of the total emission, while in C-band
 it is lower than 5\%. This effect can therefore be neglected.}
 
 \item {Differential (depth) FR: When synchrotron emission originates in a thermal plasma containing a regular magnetic field, the
 polarization plane of the radiation produced at different depths within the source is rotated over different angles by the Faraday 
 effect. The emission from deeper regions in the volume is Faraday rotated along the line of sight, and cancels out the 
 polarization from the surface layers of the volume, resulting in a decrease in the PD of the integral emission observed 
 (e.g. see \citealt{Velusamy1975} for a study of this effect in supernova remnants).  This is a wavelength-dependent
 depolarization process that is likely to require a more laminar flow than present in the WCR.
 In the case of WR~146, the large separation between the stars means that the shocks 
 in the WCR will be adiabatic, and therefore reasonably thin (although a high level of turbulence could 
 widen them). The shock width can be estimated as $\sim 0.1$ times its distance to the respective star, yielding 
 $\Delta l_1 \approx 20$~AU and $\Delta l_2 \approx 5$~AU. Typical values for the physical parameters at the shocks in the WCR are an electron 
 number density of $n_{e1} \sim 10^5$~cm$^{-3}$ and $n_{e2} \sim 10^6$~cm$^{-3}$, and a magnetic field of 
 $B_1 \sim 3$~mG and $B_2 \sim 40$~mG. Considering the simplest case of a uniform slab, we can construct a rough estimate
 of the maximum intrinsic $\mathrm{RM} \sim K_\mathrm{RM} n_e B_z \Delta l$, where the constant $K_\mathrm{RM}$ is equal to 
 $81$~cm~pc$^{-1}$~G$^{-1}$ and $B_z \sim B/\sqrt{3}$ is the component of the magnetic field in the direction of the line of sight (LOS). 
 We obtain values of $\mathrm{RM}_1 \sim 5 \times 10^3$~rad~m$^{-2}$ and $\mathrm{RM}_2 \approx 30\; \mathrm{RM}_1$.
 The depolarization factor in this case is $\xi_2(\lambda) \approx \sin{(\mathrm{RM}_2 \lambda^2)}/(\mathrm{RM}_2 \lambda^2) < 0.3\%$ 
 for $\lambda > 5$~cm, which shows that this effect could be relevant in the case of a dominant ordered magnetic field. 
 A detailed analysis of this process will be addressed in a forthcoming work, with an improved modeling of the quantities 
 $n_e$ and $B_z$ along the LOS through the shock.}
 
 \item {Internal Faraday Dispersion: In the WCR the synchrotron emitting electrons and thermal plasma coexist. In the case 
 the latter is turbulent, nearby lines of sight experience a random walk in FR. If the telescope beam encompasses many
 turbulent cells, the observed PD diminishes. This is a wavelength-dependent process. 
 Using the simple formulas given by \cite{Burn1966} for the case of a uniform slab, one has 
 $S=2 (K_\mathrm{RM} n_\mathrm{e} B_\mathrm{r})^2 (\delta L)^2 \lambda^4 \sim 10^{-2}\mathrm{RM}^2\lambda^4$, where we assumed 
 $\delta L \sim 0.1 \Delta L$ and $B_\mathrm{r} \sim B_z$. This yields $\xi_2(5\,\mathrm{cm}) \approx (1-e^{-S})/S < 0.1\%$. Once again, this 
 very simple approximation just shows that the depolarization could be strong, although a more careful analysis is needed.}
 
 \item {External Faraday Dispersion: This is similar to the internal effect above, but in this case the FR takes place in a thermal plasma
 (the stellar wind) external to the synchrotron emitting volume (the WCR). To properly account for this process, one has to solve numerically the
 value of $\mathrm{RM} = K \lambda^2 \int{ n B_z dz}$ for each emitting region in the WCR. This integral depends on both the intensity 
 and the geometry of the stellar magnetic field. In the case of a toroidal magnetic field (which is expected at large distances from 
 massive stars) supported by a surface stellar magnetic field of $B_* \sim 200$~G, the value of RM can be in the range $(3-40)\times 10^4$
 rad~m$^{-2}$, with significant dependence on the LOS for
 the particular surface element. However, this condition can be considerably relaxed if the stellar magnetic field is 
 much lower; in particular, the strength of $B_2$ could be due to a high local amplification. Other external
 Faraday screens in the interstellar medium are most likely negligible on the angular scales of interest here.}

 \item {Gradients in RM across the beam: This is similar to the dispersion effects above, but here describes variations
 in RM originating from systematic gradients within the telescope beam rather than from a random walk.
 In our case, we are looking at a curved shock that is completely unresolved within the beam of our observations.
 Then, if there is a gradient in RM along the shock, polarization vectors within the beam may cancel, leading
 to depolarization. However, one would expect that this effect does not lead to a complete depolarization of the emission
 as long as the LOS does not coincide with the axis of symmetry of the WCR (which would seldom occur and only if the orbit is seen
 edge-on at inclination $\sim 90^\circ$). Future high resolution (VLBI) polarimetric observations may be used to examine this effect. 
 Spatially resolved fractional polarization measurements of the shock could be used to discriminate between two cases:
 (i) the emission appears polarized, in which case the acceleration mechanism is likely DSA due to coherent $B$-fields, or
 (ii) the emission is still depolarized, in which case either MR or external depolarization could be responsible.} 
 
\end{enumerate}

\subsection{Source variability and possible contamination}\label{sec:variability}

The WR 146 system is a variable radio source \citep{Setia2000}.
Orbital modulation could play a significant role in its variability if the eccentricity and/or the inclination of the orbit are 
large \citep{Eichler1993}.
If the orbit is eccentric, conditions in the shocked plasma will vary, as will the medium in which the synchrotron photons
propagate. If the orbital inclination is large, this will affect the degree to which low-frequency photons emitted in the WCR
will be absorbed within the stellar wind, depending on the photon path to the observer and the orbital phase.
If the binary is seen approximately edge-on ($i\sim 90 ^\circ$), the absorption will be maximum when one of the stars is in
front of the WCR (conjunction) and minimum when they are near quadrature. However, as VLBI observations of the system have
revealed a C-shaped WCR \citep{OConnor2005}, this points to a rather small orbital inclination, 
unless those observations of the binary correspond to rather specific positions of the orbit near quadrature (if 
the system were near conjunction, the WCR would look like more circular; e.g. see Fig.~4 in \citealt{delPalacio2016}).
To explain the steep low-frequency cutoff in the spectrum seen in Fig.~\ref{fig:spectrum}, there are two other possibilities: 
(i) the effect is mostly due to the eccentricity of the orbit, in which case both observations at 325~MHz should correspond to orbital phases 
different to those of the observations at 5~GHz (some coincidence is demanded, as the observations were performed at random phases), or (ii) the 
low-frequency cutoff is related to the Razin-Tsytovich effect, which produces a cutoff at $\nu_\mathrm{RT} \sim 20 n_\mathrm{e}/B$~Hz 
(in c.g.s. units). Given that typically $n_\mathrm{e2} \sim 10^6$ in S2, this requires $B_2\sim 50$~mG. This value of $B_2$ is too small 
for MR, and therefore the cutoff is more consistent within a DSA scenario.
Nevertheless, we remark that our interpretation of the low-frequency cutoff solely in terms of free-free 
absorption is subject to the unknown orbital parameters; simultaneous observations at low ($<1$~GHz) and high ($>1$~GHz) 
frequencies would help to disentangle this ambiguity.

Moreover, it is still under debate whether the secondary in WR 146 is an O8V star or another binary \citep{Setia2000}.
Despite this uncertainty in the composition of the secondary, the very high resolution image from \cite{OConnor2005} shows signs 
of only one WCR, so it seems unlikely that our results are biased by the behavior of the secondary. Instead, the most
reasonable explanation is that there is only one dominant WCR whose variability (especially at low frequencies) is due to
geometrical modulations. Long-term monitoring is needed to confirm this.


\section{Conclusions}\label{Sec:Conclusions}

  We presented the first radio polarimetric study of a colliding-wind binary. Our VLA observations of WR 146 reveal the absence of
  linearly polarized emission down to a fractional level of 0.6\% across 1 to 8 GHz. If acceleration were due predominantly to DSA in
  strong shocks, as usually assumed, then the magnetic field component parallel to the shock front would be enhanced, and the resultant
  emission would be polarized. The absence of polarization can be explained in terms of a highly turbulent magnetic field,
  in which case turbulent magnetic reconnection would account for non-thermal particle acceleration, or depolarization effects,
  for example due to Faraday rotation in the stellar wind, or perhaps a combination of these effects. We have shown that, in principle, 
  either scenario is plausible. If turbulent MR is taking place, a high value of the magnetic field in the WCR $\sim 150$~mG is required to 
  account for the synchrotron emission. On the other hand, if DSA is the main acceleration mechanism, a magnetic field of $\approx 40$~mG is 
  sufficient to account for the synchrotron emission. However, even this relatively low magnetic field was shown to be enough to produce 
  significant FR and consequently strong depolarization. Moreover, a lower value of the magnetic field in the WCR also helps to reproduce the 
  low-frequency cutoff seen in the spectrum, as in this case the Razin-Tsytovich suppression is more effective. Our analysis seems to favor 
  a scenario where DSA and a partially ordered $B$-field are responsible for the synchrotron emission, while various depolarization 
  processes account for the very low PD in the radiation. Future high resolution observations of the shock could be used to reveal polarized
  emission, where the angular scale at which the shock appears polarized would be an important diagnostic discriminant between MR and DSA.
  An RM detection would also help to distinguish between MR and DSA by providing constraints on magnetic field strengths, and thus checking
  if MR remains plausible. In a forthcoming work we will investigate the depolarization effects in more detail,
  and study if the detection of polarization remains elusive in other particle-accelerating colliding-wind binaries.


\begin{acknowledgements}
      PB and GER are members of the CIC, CONICET.
      GER acknowledges support by the Spanish Ministerio de Econom\'{i}a y Competitividad (MINECO/FEDER, UE) under
      grant AYA2016-76012-C3-1-P and CONICET grant PIP 0338.
      The National Radio Astronomy Observatory is a facility of the National Science Foundation operated
      under cooperative agreement by Associated Universities, Inc.
\end{acknowledgements}

\bibliographystyle{aa} 
\bibliography{biblio} 

  \section*{Appendix}

  Tab.~\ref{tab:data} presents our VLA flux density measurements for WR~146, as displayed in Fig.~\ref{fig:fit_alpha}.
  
\begin{table}[]
\centering
\begin{tabular}{cccc}

\hline

\textbf{$\nu$ [GHz]} & \textbf{$S_\nu$ [mJy]} & \textbf{$\sigma_{S_\nu}$ [mJy]}\\ \hline
1.040 & 95 & 0.45 \\ 
1.296 & 86.4 & 0.34 \\ 
1.360 & 84.7 & 0.21 \\ 
1.424 & 82.7 & 0.18 \\ 
1.488 & 81.6 & 0.19 \\  
1.680 & 73.5 & 0.22 \\  
1.744 & 73.6 & 0.2 \\  
1.808 & 72.9 & 0.19 \\  
1.872 & 72 & 0.19 \\  
1.936 & 70.3 & 0.29 \\  
2.000 & 68.4 & 0.25 \\  

\hline

4.167 & 45.2 & 0.45 \\  
4.295 & 44 & 0.37 \\  
4.423 & 44.4 & 0.39 \\  
4.551 & 42.3 & 0.33 \\  
4.679 & 42 & 0.33 \\  
4.807 & 41.6 & 0.38 \\  
4.935 & 40.3 & 0.37 \\  
5.063 & 40.2 & 0.39 \\  
5.191 & 39.1 & 0.37 \\  
5.319 & 38.3 & 0.32 \\  
5.447 & 38.4 & 0.35 \\  
5.575 & 37.7 & 0.39 \\  
5.703 & 37 & 0.33 \\  
5.831 & 36.3 & 0.37 \\  
5.959 & 36.9 & 0.43 \\  
6.039 & 36.1 & 0.43 \\  
6.295 & 34.3 & 0.37 \\  
6.551 & 33.3 & 0.33 \\  
6.679 & 32.6 & 0.34 \\  
6.807 & 32.5 & 0.35 \\  
6.935 & 33.7 & 0.39 \\  
7.063 & 32.1 & 0.36 \\  
7.191 & 32.3 & 0.35 \\  
7.319 & 32.4 & 0.35 \\  
7.447 & 32.9 & 0.38 \\  
7.575 & 30.7 & 0.32 \\  
7.703 & 30.6 & 0.3 \\  
7.831 & 30.2 & 0.32 \\ 
\hline
\end{tabular}
\caption{
VLA flux densities for WR~146 measured across L-band and C-band within
64~MHz and 128~MHz spectral windows, respectively.
}
\label{tab:data}
\end{table}

\end{document}